\documentclass[preprint,5p]{elsarticle}
\usepackage{amssymb}
\usepackage{caption}
\usepackage{subcaption}
\usepackage{amsmath}
\usepackage{xcolor}
\usepackage{lipsum}
\usepackage{graphicx}
 \usepackage[normalem]{ulem} 

\journal{Surfaces and Interfaces}

\begin{document}

\begin{frontmatter}



  \title{\textbf{Permeation of hydrogen across graphdiyne: molecular dynamics vs. quantum simulations and role of membrane motion}}


\author{Mateo Rodr{\'i}guez}
\author{Jos{\'e} Campos-Mart\'{\i}nez}
\author{Marta I. Hern\'{a}ndez\corref{cor1}}\ead{marta@iff.csic.es}

\cortext[cor1]{correspondence to}
\address{{Instituto de Física Fundamental, Consejo Superior de Investigaciones Cient{\'\i}ficas (IFF-CSIC)},
            {Serrano 123}, 
            {Madrid},
            {E-28006}, 
            {Spain}}

\begin{abstract}
 Previous research based on electronic structure calculations and molecular
dynamics (MD) simulations have demonstrated that graphdiyne (GDY) is a
very suitable two-dimensional membrane for the separation of
small molecules in a gas mixture of different species.
However, quantum effects may play a role in the dynamics of these
permeation processes when light molecules are the ones involved
in the crossing of the GDY subnanometric pores.
In this work we report rigorous quantum-mechanical calculations
together with equivalent MD simulations of the transport of H$_2$
molecules through a static GDY membrane, as a case study for the validity
of the application to these problems of classical dynamics.
The force fields employed are based on an improved Lennard-Jones formulation,
with parameters optimized by means of accurate {\em ab initio} calculations.
It is found that, although quantum effects are still significant at
the temperatures of interest (between 250 and 350 K),
MD simulations are able to reasonably reproduce the dependence
of the quantum permeances with the temperature.
Moreover, MD permeances computed with quantum corrections through
Feynman-Hibbs effective potentials provide a lower bound
to quantum permeances, while the pure classical counterpart gives
an upper bound, thus leading to a well delimited range of confidence
of the permeation results.
Furthermore, within MD simulations it is possible to incorporate 
the thermal motion of the GDY layer and in this situation it is observed an
enhancement of the 
permeances with respect to the fixed membrane case,
due to a significant reduction of the permeation barriers when
the GDY atoms are allowed to vibrate.
It seems apparent therefore, that modeling the membrane motion is
crucial to provide reliable simulations of the gas transport features.
\end{abstract}

%



\end{frontmatter}




\section{Introduction}

Membrane technologies are of remarkable importance for applications of separation
of atoms or molecules from gaseous or liquid mixtures\cite{membranes-separa-koros-1993,Bernardo-IJHE-2020,Azizi-JWPE-2025}.  
Since membrane permeances are higher for thinner membranes, the ultimate membrane would be a one-atom thick membrane, the celebrated 2D materials\cite{Wang-NatNanotech-2017}. 
In these materials, the existence of nanopores is crucial to provide the required confinement/selectivity for the different species and ultimately for isotopic separation.  
Graphynes (GY)\cite{jia_synthesis_2017,sakamoto_accelerating_2019,kang_graphyne_2019,gr2-review-22} 
are a very popular group of nanoporous carbon-based 2D materials, with several allotropes known as
$\alpha,\beta,\gamma-$GY,\cite{sakamoto_accelerating_2019,li_structural_2020} being a bit more stable the $\gamma$ family\cite{graphy-rev-energy-25}. 
This variety is well described by benzene type hexagons with vertex joined by triple acetylenic bonds, leading to triangular nanopores regularly disposed.
Thus, they can be named as graph-$N$-ynes, with $N$ the number of acetylene type molecules between the rings.
Graphdiyne (GDY)
\cite{li_architecture_2010,jacs-synthesis-gr2-17,gr2-synthe-propert-17} and graphtetrayne\cite{gao_architecture_2018,2dn-GRT-syntehsis-21} have been already synthesized, and it is GDY one of the materials most widely used\cite{gr2-review-22,Ren-S&I-2025} and, therefore, almost routinely synthesized in the lab, in particular as an efficient membrane for separation of different species\cite{zhou_gas_2022,Li_NatWater_2025,xu_molecular_2020,Cranford-Nanoscale-2012}.

In this context, most of the theoretical studies on the separation of different gas species using 2D membranes have resorted to electronic structure calculations\cite{jiao_graphdiyne_2011,Zhang-JPhysD-2013,zhang_tunable_2012,Tian-ApplNanoMat-2021,Mahnaee-S&I-2025,Liu-SurfInt-2024,Guo-S&I-2024} or classical molecular dynamics (MD) simulations \cite{Sun-Langmuir-2014,Cranford-Nanoscale-2012,Zhao-CJCP-2012,Apriliyanto-JPCC-2018,Sun-JPCL-2019,Liu-Membranes-2020,Azizi-SciRep-2021,Tian-ApplNanoMat-2021,Liu-SurfInt-2024,Guo-S&I-2024}, leading to useful trends and predictions of permeances and selectivities of different kinds of molecular species.
However, quantum effects (such as tunneling or quantization of the transition state at the pore center) can be important in the dynamics of these processes, specially those involving light molecules and/or low temperatures.
Therefore, it would be very interesting to examine the performance of MD simulations in comparison with quantum dynamics calculations in conditions where both classical and quantum models were as equivalent as possible, since we are not aware of research relevant to this issue in the literature. 
Most of the quantum approaches adopted so far are somehow approximate, such as the use of Feynman-Hibbs effective potentials\cite{FH-book,kumar-PRL-2005}, Ring Polymer Molecular Dynamics methods\cite{Craig_JCP:2004,rpmd-isotop-he-gr2-2021} or quantum models of reduced dimensionality\cite{Schrier_JPCL:2010,Bartolomei_JPCL:2014,Hernandez-JPCA-2015,Qu-SciRep-2017}.
Some of us have contributed to the development of more accurate quantum methods for the transmission of molecules across 2D membranes, such as three-dimensional time-dependent wave packet (TDWP) method to compute transmission probabilities \cite{Gijon-JPCC-2017} and, more recently, permeances at different temperatures\cite{Arroyo-PCCP-2022}.
In the latter work, the permeation process is modeled as a molecule that can approach a rigid membrane from any incidence direction, in this way resembling the interaction of a dilute gas with a static membrane, which can be typically addressed by means of MD simulations.

\begin{figure}[h!]
\centering
\includegraphics[width=9.cm,angle=0.]{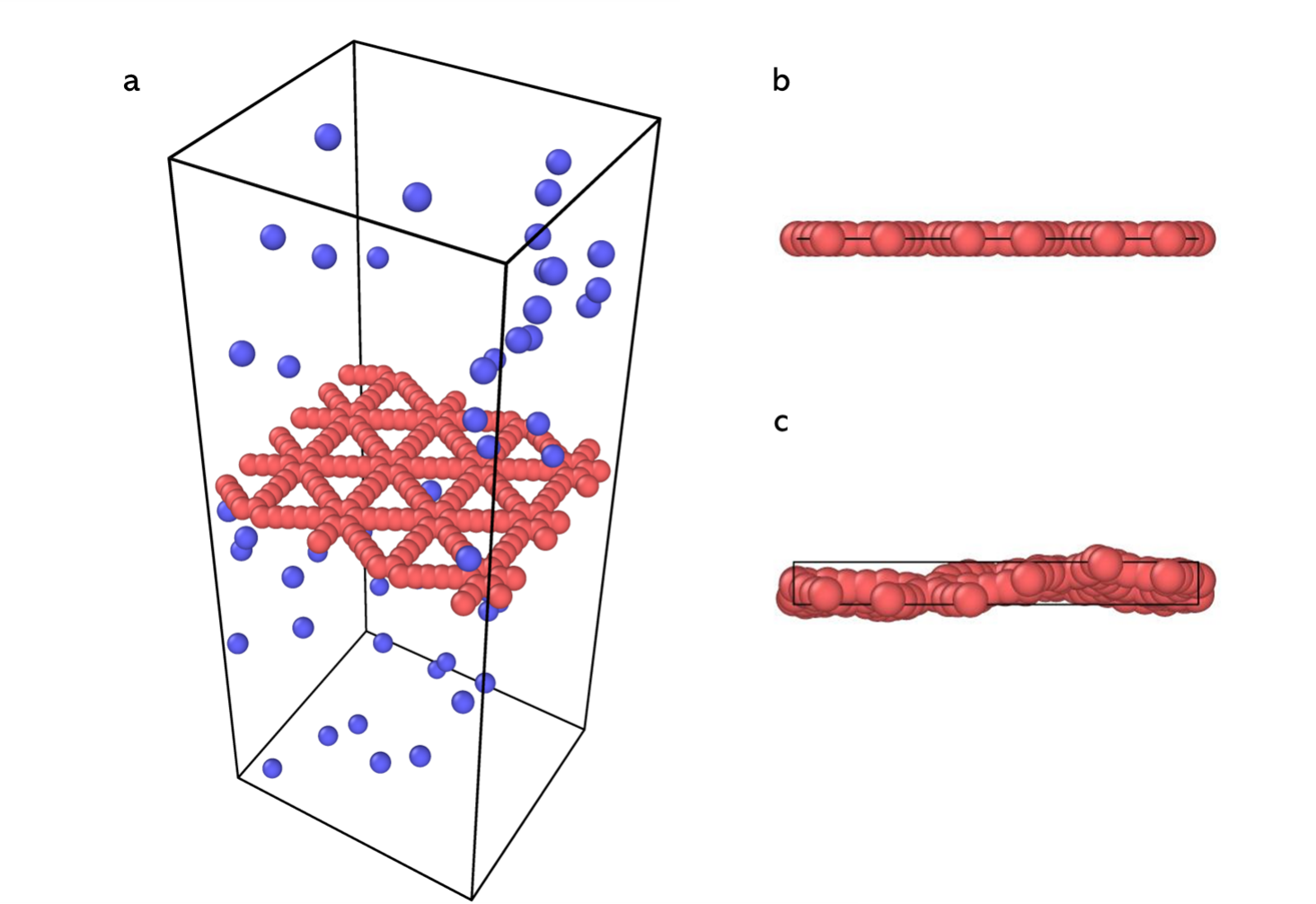}
\caption[] {(a) Simulation box for MD simulations, with H$_2$ molecules represented as point particles. Note that, to save space,  its height and number of molecules is half of the box actually used in the MD calculations. (b) and (c) Lateral view of the GDY membrane for fixed- and deformable-membrane simulations, respectively.}
\label{Fig_simbox}
\end{figure}

In this work we have extended previous TDWP calculations of permeances of a static GDY membrane for the transport of hydrogen\cite{Arroyo-PCCP-2022} to higher temperatures (250-350 K) to compare with equivalent MD simulations at the same temperatures. 
In both classical and quantum approaches, we describe the H$_2$ molecules as pseudo-atoms (point particles with the mass of H$_2$), employing interaction potentials (force fields) that correspond to averages over the different orientations of these molecules\cite{Arroyo-PCCP-2022,Ortiz-PCCP-2019}.
We have also carried out MD simulations with Feynman-Hibbs effective potentials in order to examine the ability of these modifications in approaching the quantum results. 
A scheme of the setup for the MD simulations is presented in Fig.~\ref{Fig_simbox}.a.
As in the quantum calculations, the membrane is fixed to its planar structure throughout the MD simulations (Fig.~\ref{Fig_simbox}.b). 
Quantum simulations describing the motion of the membrane  are not feasible due to the enormous number of degrees of freedom involved, except in some specific approximate treatments\cite{Jiang-JPCL-2021}.

As a second goal of this work, we have examined the role of the motion of the atoms of GDY in the transport dynamics of H$_2$ molecules, within the MD framework.
As in free-standing graphene\cite{DeAndres-PRB-2012,Xu-NatComm-2014}, one can expect that a GDY sheet may also exhibit large amplitude undulations or ripples due to thermal fluctuations or interactions with molecules, indeed, as can be seen in Fig.~\ref{Fig_simbox}.c.
This motion may affect the membrane permeances for the transport of the gas molecules, an aspect that, however, has been scarcely studied\cite{Bucior-JPCC-2012,Azizi-SciRep-2021}.
Here we find that the permeances obtained by allowing the motion of the GDY membrane are significantly larger than those corresponding to the fixed membrane approximation and analyze the rationale of this important role of the membrane.

The paper is organized as follows. 
In Section 2 we present a detailed account of the computational methods employed: force fields, quantum and MD calculations.
Results are shown in Section 3, with discussion focused on the comparison of classical and quantum permeances (fixed membrane) as well as the role of the membrane motion in the modification of the classical permeances.
Finally, concluding remarks are given is Section 4. 
Additional clarifications and graphics are provided in a Supplementary Information (SI) document.

\section{Computational methods}

In this study the hydrogen molecules are treated as point-particles (i.e., only their translational degrees of freedom are considered), an approach that is adequate to describe 
the ground state of the molecules when they are subjected to nearly isotropic interactions\cite{Silvera-JCP-1978,Calvo-JCP-2016}. Within this approximation, we first present the force field used, followed by details of the quantum dynamics calculations (frozen membrane) and of the molecular dynamics simulations (both for frozen and mobile membrane). 

\subsection{Modeling of the force field}\label{forcefield}
The interactions between pairs of H$_2$ molecules as well as between these molecules and the carbon atoms of the GDY membrane have been represented employing the Improved Lennard-Jones (ILJ) potential of Pirani {\em et al}\cite{ILJ}. 
This pairwise potential adds an extra parameter with respect to the widely used Lennard-Jones function, providing in this way a more realistic representation of both the size repulsion and the long-range attraction of the partners and therefore producing successful comparisons with scattering data, transport properties, etc. (see Ref.\cite{ILJ} and cites therein).
This function takes the form

\begin{equation}
 V_{\rm ILJ}(R)  \!= \!
  \varepsilon  \!\left[\frac{\gamma}{n(R) \!- \!\gamma}
  \! \left (\frac{R_m}{R} \right)^{n(R)}
    \! -  \!\frac{n(R)}{n(R) \!-\gamma} \!
   \!  \left(\frac{R_m}{R}\right)^{\gamma} \right]
\label{EqILJ}
\end{equation}

\noindent
where $R_m$ is the equilibrium distance, $\varepsilon$ the well depth, $\gamma$ is related to the long range behavior of the interaction and $n(R)=\alpha \frac{R}{R_m}+\beta$, with $\alpha=4$ and $\beta$ being a parameter that defines the shape and stiffness of the potential. 
The typical Lennard-Jones formula is recovered for  $\gamma$=6, $\alpha$=0 and $\beta$=12. 
This ILJ potential has been implemented as a new pair style within LAMMPS\cite{lammps}, the code used to run the classical simulations ({\em lj/pirani} therein, see also Ref.\cite{Rodriguez-DCSIC-2025}). 
Values of these parameters for the H$_2$-C and H$_2$-H$_2$ interactions have been determined\cite{Arroyo-PCCP-2022,Ortiz-PCCP-2019} by comparisons with accurate {\em ab initio} calculations\cite{Bartolomei-Carbon-2015,Patkowski-JCP-2008,Ortiz-PCCP-2019} and they are provided in Table~S1 of the Supporting Information (SI) document. 
In the molecular dynamics simulations, these pair interactions operate up to a cutoff distance of 8 \r{A}.

For the molecular dynamics studies where the membrane is considered to be deformable, we have added the AIREBO potential\cite{Stuart-JCP-2000} (cutoff distance of 2.5 \r{A}) to model the interactions among the carbon atoms of GDY.

 \begin{figure}[h!]
 \centering
\includegraphics[width=8.5cm,angle=0.]{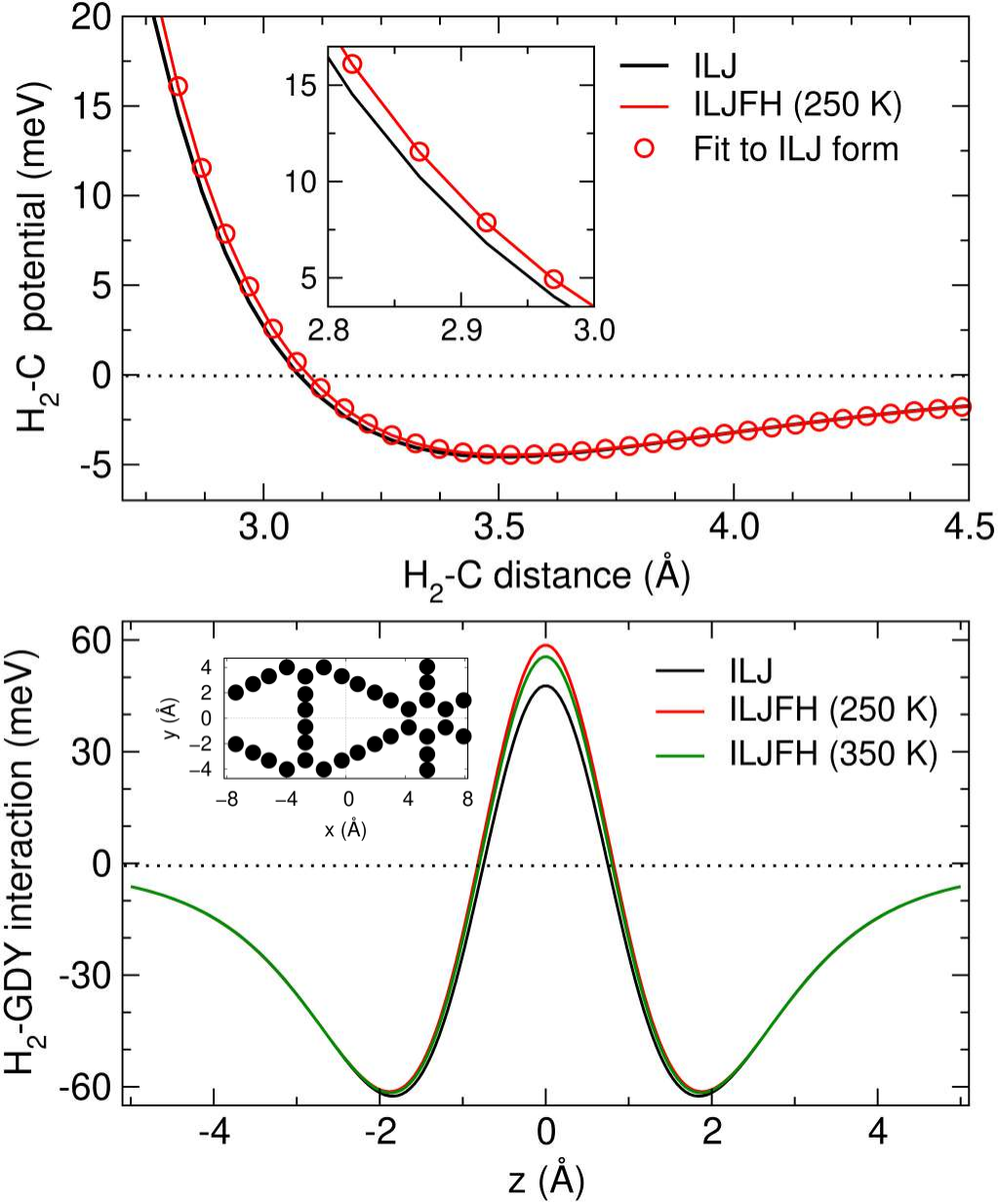}
\caption[] {Upper panel: H$_2$-C pair potential (meV) vs. intermolecular distance (\AA). 
The ILJ potential (Eq.~\ref{EqILJ}) is shown in black line, while red lines and red open circles are used for the ILJFH at 250 K (Eq.~\ref{EqFH2}) and its fit to a ILJ formula, respectively.
The inset presents a zoom of the region below 3 \AA.
Lower panel: Profile of the H$_2$-GDY interaction potential (meV) along $z$ (\AA), the direction perpendicular to the membrane plane, crossing the center of a pore. 
In black line, bare ILJ force field, whereas the ILJFH potentials are shown in red and 
green lines, for 250 and 350 K, respectively.
The inset shows the GDY unit cell, with the center of a pore at $(x,y,z)=(0,0,0)$.
}
 \label{Fig-pots}
 \end{figure}

In order to achieve estimations of quantum effects at the level of the classical simulations, we have also run calculations using Feynman-Hibbs (FH) effective potentials\cite{FH-book}, which correct the force fields with terms depending on the masses of the partners and the temperature.
This formulation has been widely applied to the study of the adsorption and sieving of small molecules to different substrates\cite{kumar-PRL-2005,Contescu-PRL-13,Rodriguez-JPCA-2016}.
Specifically, the second-order FH correction to the ILJ potential (ILJFH in the following) reads

\begin{equation}
V_{\rm ILJFH}(R)  =  V_{\rm ILJ}(R) \,
 + \, \frac{\hbar^2 \beta}{24 \mu} 
\left( \, V_{\rm ILJ}''(R) + \frac{2}{R}V_{\rm ILJ}'(R) \, \right), 
\label{EqFH2}
\end{equation}

\noindent
where $\beta=1/K_BT$, $\mu$ is the reduced mass of the interacting particles and 
$V_{\rm ILJ}'(R)$ and $V_{\rm ILJ}''(R)$ are the first and second derivatives of the ILJ function.  
For the different temperatures studied, we have found that the ILJFH potentials can be accurately fitted to ILJ formulas, so that they are introduced into the LAMMPS code in their corresponding ILJ forms.
Values of the corresponding parameters are are given in Table~S2 of the SI for both the H$_2$-C and H$_2$-H$_2$ interactions. 

In the upper panel of Fig.~\ref{Fig-pots} we present the H$_2$-C ILJ potential, the corresponding ILJFH one at 250 K as well as the fit of the latter to an ILJ expression.
It is seen that the FH correction is small except in the short range, where the interaction becomes more repulsive when the correction is added (see inset), a feature that is relevant for the H$_2$-GDY interaction in regions close to the pores of the membrane.
The lower panel of Fig.~\ref{Fig-pots} shows ILJ and ILJFH  H$_2$-GDY interaction potentials in a direction perpendicular to the layer and passing through the center of a pore.
The GDY rectangular unit cell is shown in the inset of that panel, while the 
positions of their carbon atoms are reported in Table S3 of the SI.
The H$_2$-GDY potentials have been obtained by summing all needed H$_2$-C pairwise contributions until convergence is reached.
Compared to the ILJ bare potentials, the FH effective potentials lead to higher permeation barriers: the barrier height rises from 48 meV for the bare potential 
to 55 and 59 meV for the ILJFH potentials at 350 and 250 K, respectively.
In this way, the FH correction approximately incorporates quantum effects such as the zero point energy of the in-pore degrees of freedom that makes the quantum permeation exhibit a higher energy threshold than the classical one\cite{Hernandez-JPCA-2015,Gijon-JPCC-2017,Arroyo-PCCP-2022}.
Therefore the permeance of the H$_2$ gas will decrease when these corrections are added to the ILJ potentials, as will be seen in the next Section.

\subsection{Time-dependent wave packet (TDWP) quantum calculations}

The method for the calculation of permeances within a quantum-mechanical framework 
has been provided in Ref.\cite{Arroyo-PCCP-2022}, so a brief summary is just presented 
here.
Having the flux $Z$ of the molecules hitting the membrane, defined as $Z=\rho v_z$ ($\rho$ being the number density
of the molecular gas and $v_z$ the perpendicular velocity of the molecule) and the probability that these molecules
actually cross that surface, ${\cal P}$, we can obtain the thermal flux $J(T)$ as the thermal average of the product of $Z$
and the mentioned probabilities.
The permeance $S$ of the membrane is $J(T)$ divided by the pressure, $P$, which, considering that $\rho=P/K_B T$,
can be written as

\begin{equation}
S(T)  =  \frac{\langle v_z   {\cal P} \rangle_T}{K_BT} 
     =  \frac{1}{K_BT} \int_{-\infty}^{\infty} d{\bf v} \, v_z \, f({\bf v},T) \, {\cal P}({\bf v}), 
\end{equation}

\noindent
where $f({\bf v},T)$ is the Maxwell-Boltzmann distribution of velocities. 

The probabilities ${\cal P}({\bf v})$ are obtained by means of TDWP calculations which simulate the impact (and eventual crossing) of a H$_2$ molecule on the membrane.
The wavepackets $\psi(x,y,z,t)$ are represented on a grid where the periodicity of the membrane is exploited\cite{Yinnon-CPL-83} by matching the size of the $(x,y)$ box to that of the unit cell (inset of the lower panel of Fig.~\ref{Fig-pots} and Table~S1 of the SI).
They are propagated subjected to the time-dependent Schr{\"o}dinger equation, using the split operator method\cite{SplitOperator} and the fast Fourier transform between the position and momentum spaces.
At initial time, they are given as

\vspace{-.2cm}

\begin{equation}
  \psi (x,y,z,t=0) = G(z;z_0,v_{z0}) \,
  \frac{ \exp \left[i \;m (v_x x+v_yy)/\hbar\right] } {(\Delta_x \Delta_y)^{1/2}}, 
  \label{initialwp}
\end{equation}

\noindent
where $G(z;z_0,v_{z0})$ is a Gaussian wavepacket\cite{Arroyo-PCCP-2022} involving a distribution of perpendicular velocities $v_z$ around a central velocity $v_{z0}$.
The other factor is a plane wave with well-defined parallel velocity $(v_x,v_y)$, 
$m$ being the mass of H$_2$ and $(\Delta_x,\Delta_y)$, the dimensions of the unit cell.
The values of $(v_x,v_y)$ are restricted so that the plane wave is commensurate with the lattice, namely,

\begin{equation}
(v_{l_x},v_{l_y})= \frac{2\pi\hbar}{m}\left( \frac{l_x}{\Delta_x},\frac{l_y}{\Delta_y} \right),    
\label{Eq-parvel}
\end{equation}

\noindent
where $(l_x,l_y)$ are integer numbers.
Details of the calculations are given in the SI.

\subsection{Molecular dynamics simulations}

MD simulations have been carried out using the force fields described in subsection~\ref{forcefield} and employing LAMMPS, the Large-scale Atomic/Molecular Massively Parallel  Simulator \cite{lammps}. 
The calculations were performed within the canonical (NVT) ensemble and considering the GDY membrane either fixed at its planar structure or allowing it to evolve in time, in the latter case modeling the C-C interactions with the previously mentioned AIREBO force field.

The simulation box has a size of approximately $3.2\times2.8\times18.0$ nm$^3$.
The box size along ($x,y$) equals the size of 2$\times$3 GDY rectangular unit cells (inset of Fig.~\ref{Fig-pots}) so that the membrane, initially coincident with the $z=0$ plane, contains a total of 216 carbon atoms. In the case of the mobile membrane, the position of one carbon atom has been fixed throughout the simulations to avoid vertical displacement of the membrane.
Periodic boundary conditions are used in the $x$ and $y$ directions, while reflective boundary conditions are taken at the limits of the box in the $z$ direction. 
The system is shown in Fig.~\ref{Fig_simbox} where, for the sake of saving space, the size along $z$ has been reduced to 9 nm, maintaining the density of hydrogen's constant. 
The equations of motion were solved using an integration time step of 0.1 fs and 
a Nos{\'e}-Hoover thermostat has been chosen to resemble the NVT ensemble at different temperatures, with a relaxation constant of 10 fs.
For every MD simulation an equilibration period of 1 ns has been taken, followed by 4 ns of data acquisition.

We consider 100 molecules of hydrogen, initially 50 on each side of the membrane, corresponding to nominal pressures between 20-29 bar for the studied range of temperatures, between 250-350 K. 
In order to compute fluxes and permeances, we have adopted the method of Ref.\cite{Sun-Langmuir-2014}, where the molecules are in equilibrium in both sides of the membrane and the flux is computed by counting crossings in either direction.
For a dilute gas, it is expected that this flux is twice the flux in a non-equilibrium system where all the molecules are initially on one side of the membrane. 
In the two-sided equilibrium procedure, the number of crossings vs. time approximately draws a straight line from which the flux (dividing the slope by the area and by two, due to the two directions of crossings) and hence permeance (flux divided by pressure) are easily estimated, in comparison with the non-equilibrium simulation, where the dependence of the number of crossings with time and subsequent calculation of permeances is more complex due to the variation with time of the pressure drop between the two chambers separated by the membrane.
Also as in Ref.\cite{Sun-Langmuir-2014}, the simulation box is divided into an "adsorption region" from $-z_{ads}$ to $z_{ads}$, where the molecules are assumed to be adsorbed, and two "bulk regions", [-$z_{max}:-z_{ads}$]  and [$z_{max}:z_{ads}$], where the molecules are in the gas phase.
As in previous studies \cite{Sun-Langmuir-2014,Cranford-Nanoscale-2012}, we count the molecules that cross the membrane from one bulk region to the other one.

As expected from a previous quantum study\cite{Arroyo-PCCP-2022}, permeances in this system are quite small, therefore, a 5 ns long simulation usually produces a rather small number of crossings. 
In order to overcome statistical uncertainty, for each force field and temperature, the permeance has been obtained from a large number of 5-ns simulations with different initial conditions, from which an average result is obtained. 
This procedure is described in detail in the SI. 

Finally, some test calculations for different pressures or for an enlarged membrane area have been performed, in order to validate the generality of our results.

\section{Results and discussion}

We start this section with a presentation of some features of the quantum-mechanical (TDWP) as well as MD simulations.
Next, we focus onto the comparison of the MD permeances with the quantum mechanical ones for case of the fixed membrane model.
The section ends with a MD study and discussion of the role of the motion of the membrane on the transport of the H$_2$ molecules across GDY.

\begin{figure}[h!]
\centering
\includegraphics[width=6.4cm,angle=-90.]{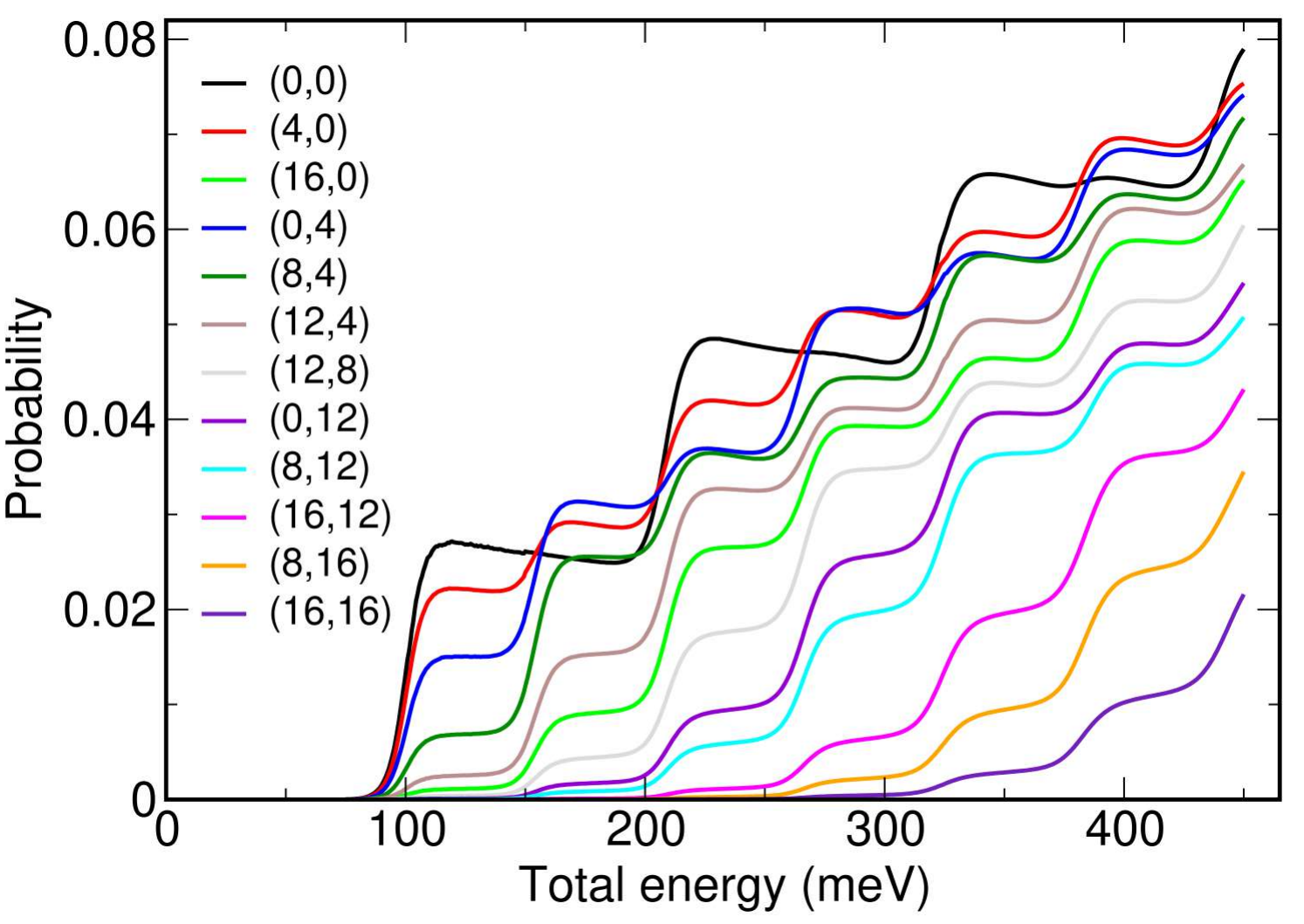}
\caption[] {Quantum probabilities for the transmission of H$_2$ across GDY, as functions of the total incident energy (meV) and for different incident parallel velocities, $(v_{l_x},v_{l_y})$ (Eq.~\ref{Eq-parvel}), labeled by $(l_x,l_y)$ in the legend.}
\label{Fig-Probs}
\end{figure}

In Fig.~\ref{Fig-Probs} we present some of the computed quantum probabilities. 
For each incident parallel velocity, $(v_{l_x},v_{l_y})$, the probabilities are presented, labeled with $l_x,l_y$, as functions of the total energy, i.e., ${\cal P}_{l_x,l_y}(E)$ with $E=m(v_{l_x}^2+v_{l_y}^2+v_z^2)/2$.
It is seen that the probabilities become smaller as the parallel velocities increase, in other words, for a given total energy, the probabilities are higher for larger values of the perpendicular component, $v_z$. 
It is worth emphasizing the quantum nature of these probabilities: their stair-case shape
as functions of the energy is due to the quantization (energy levels) of the transition state at the pore center, as discussed in detail elsewhere\cite{Hernandez-JPCA-2015,Gijon-JPCC-2017,Arroyo-PCCP-2022}.
In particular, it is noted that the probabilities rise at about 100 meV (first "step" of the staircase), an energy much higher than the value of the classical barrier (48 meV); this is explained by the ground energy level, or zero-point energy, of the transition state.
It is remarkable that this quantum structure is clearly kept up to the highest energies considered, a result that is due to the light mass of hydrogen as well as its confinement within the pore.

\begin{figure}[h]
\centering
\includegraphics[width=9.2cm,angle=0.]{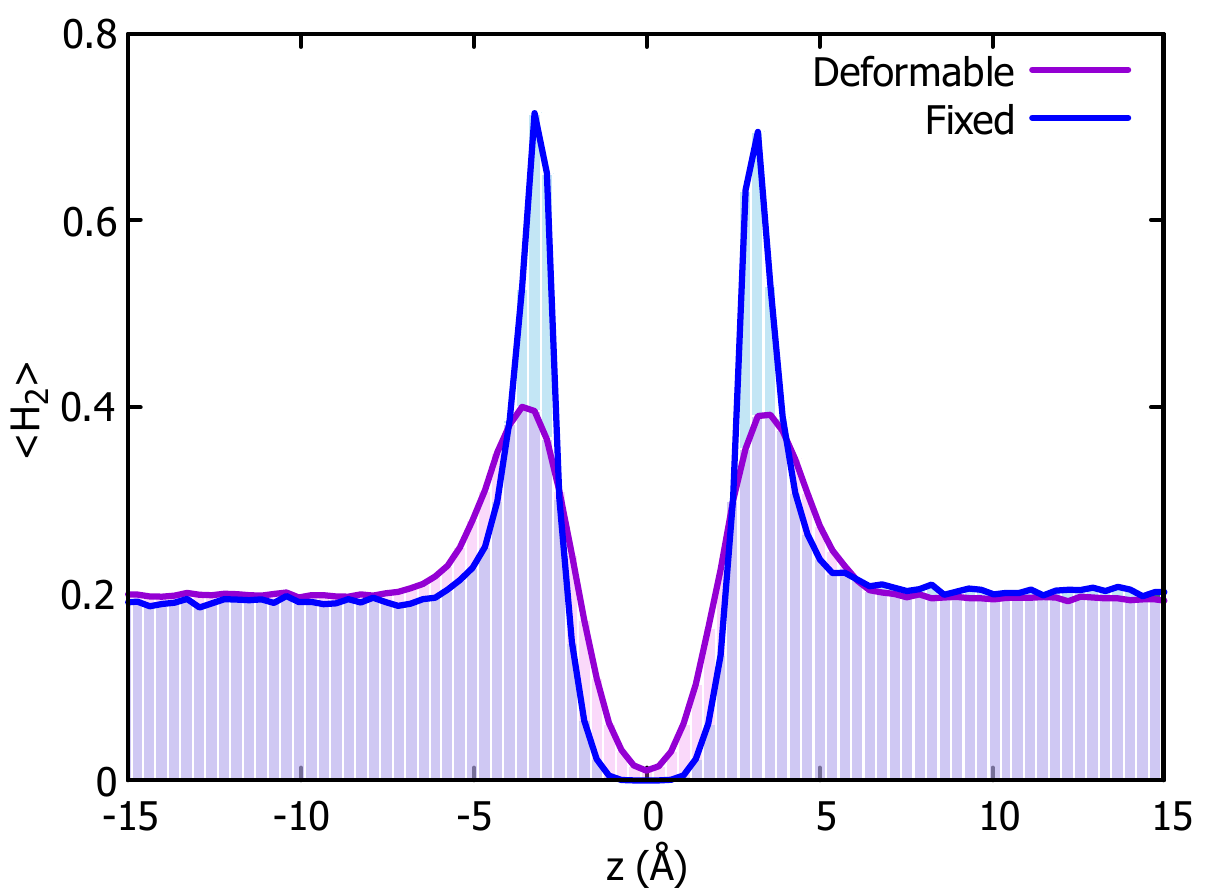}
\caption[] {Hydrogen molecules distribution at T = 300 K for MD simulations considering ILJ+AIREBO interactions and both fixed and deformable membranes.}
\label{Fig4}
\end{figure}

Moving to the MD calculations, Fig.~\ref{Fig4} shows the distribution of the hydrogen molecules along $z$, the direction perpendicular to the membrane, for both the fixed (ILJ) and deformable (ILJ+AIREBO) membrane simulations at a temperature of 300 K. 
It can be seen that the adsorption and bulk regions can be clearly distinguished:
in the adsorption region, the distribution exhibits a minimum at $z=0$ (where the fixed membrane is) and two maxima at about $z=\pm$ 3.5 \AA \, whereas it becomes approximately constant for larger values of $|z|$, in the bulk region.
Some differences between the distributions of the simulations with fixed and deformable membranes can be noticed: the molecular population around $z=0$ is almost zero for the fixed membrane, but is non-negligible in the case of the deformable membrane. 
Also, the adsorption region extends towards larger values of $|z|$ in the deformable membrane simulations.
In fact, we have defined the limits of the adsorption region $([-z_{ads},+z_{ads}])$
to be $z_{ads}=5.0$ and 6.0 \r{A} for the rigid and deformable surfaces, respectively. Defining the adsorption region to be larger than these values does not influence the results.
These differences are due to the motion of the membrane atoms which involve oscillations where some regions of the membrane displace either towards positive or negative values of $z$ (see Fig.~\ref{Fig_simbox}(c) for instance).
As a consequence, some H$_2$ molecules more easily reach distances close to $z=0$. 
Also, the distribution of the adsorbed layer is adjusted to the oscillations of the membrane, resulting to a wider distribution.
Nevertheless, the number of molecules in the adsorption region is very similar for the fixed and deformable membrane simulations: about 7\% and 8\%, respectively.

\begin{figure}[h!]
\centering
\includegraphics[width=9.cm,angle=0.]{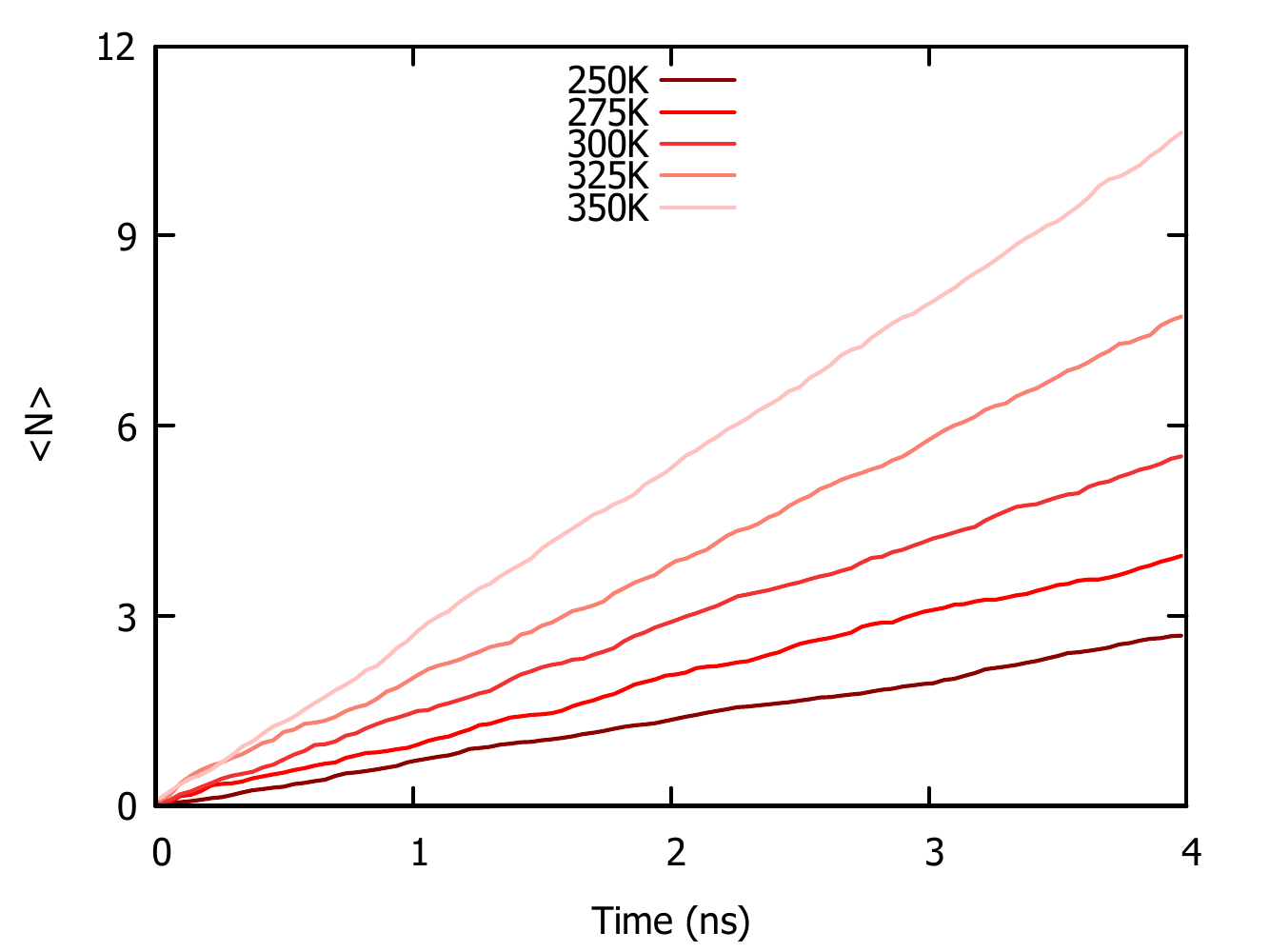}
\caption[] {Average number of crossings through fixed GDY as a function of time for each of the simulated temperatures, considering ILJFH interactions.}
\label{Fig-crossings}
\end{figure}

Fig.~\ref{Fig-crossings} presents the number of crossings across the membrane as a function of time for the different temperatures considered, specifically, for the simulations with a fixed membrane and the ILJFH force field.
As indicated in the previous section and detailed in the SI, the crossings shown in the figure are in fact an average value of the crossings observed in a large number  of simulations (between 40 and 175), carried out in parallel with different initial conditions.
It can be seen that, thanks to this procedure, the crossings (which are very few indeed) closely follow a linear behavior with time, from which the flux and the permeance can be easily estimated. 
Fig.~S1 of the SI shows an study of the convergence of this procedure with the number of simulations.
Analogous figures of the number of crossings vs. time, for the cases of the ILJ interactions (fixed and mobile membrane) are presented in Fig.~S2 of the SI.
A test calculation considering a unique long-time simulation (50 ns) is provided in Fig.~S3 of the SI, for comparison.

\begin{figure}[h!]
\centering
\includegraphics[width=9.cm,angle=0.]{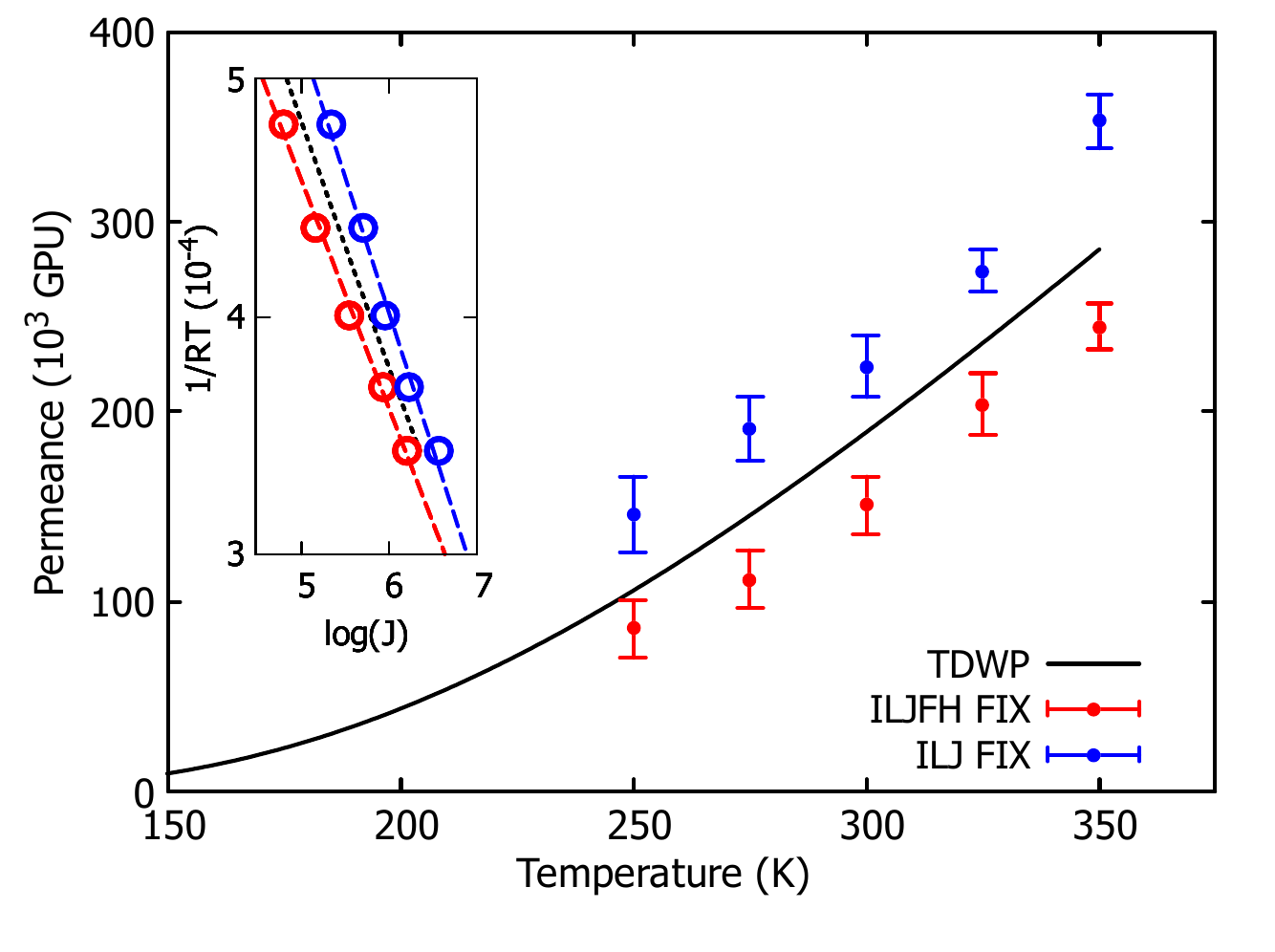}
\caption[] {Permeances (in $10^3$ GPU, 1 GPU=3.35$\cdot 10^{-10}$ mol/(m$^2$sPa)) vs. temperature (in K) of a static GDY membrane for the flow of molecular hydrogen, as obtained from MD simulations (ILJ and ILJFH interactions in blue and red points) and from quantum-mechanical TDWP calculations (black line).
The inset shows $1/(RT)$ vs. the natural logarithm of the flux $J$ (in SI units) for the same simulations (empty circles), while the dashed lines correspond to fits to Eq.~\ref{eq_arrh}, from which the activation energies are estimated.
}
\label{Fig6}
\end{figure}

Classical MD (ILJ and ILJFH interactions) and quantum (TDWP and ILJ force field) permeances are reported in Fig.~\ref{Fig6} as a function of temperature,  for the case of a fixed GDY membrane.
Numerical values of those permeances are presented in Table~\ref{Tab1}.
It can be seen that the classical ILJ permeances are larger than the quantum permeances.
The fact that the classical permeances overestimate the quantum ones is an expected result, since the quantum transport becomes limited by the energy levels structure of the transition state, leading to an energy threshold for the transmission probabilities larger than the potential barrier as well as to their stair-like structure (Fig.~\ref{Fig-Probs}).
Also, notice that the relative errors of the classical with respect to the quantum ones, 
given in Table~\ref{Tab1}, tend to decrease with temperature, in agreement with the expectation that the classical observable should approach the quantum one as temperature increases.
Classical permeances obtained with the ILJFH force fields, on the other hand, become lower than the quantum permeances.
They are closer to the quantum permeances than the ILJ ones, as their relative errors 
(Table~\ref{Tab1}) are smaller.
It is somewhat surprising the large variation of the ILJFH permeances with respect to the ILJ ones, taking into account the relatively small Feynman-Hibbs corrections involved (e.g., see Fig.~\ref{Fig-pots}), indicating the sensitivity of the simulations to fine details of the force fields employed.

\begin{table}[h!]
\centering
\begin{tabular}{|c|rrrrr|}
\hline
\multicolumn{1}{|c|}{}                & \multicolumn{5}{c|}{Permeance ($10^3$ GPU)}                          \\ \hline
\multicolumn{1}{|c|}{T (K)}    & \multicolumn{1}{c|}{TDWP } & \multicolumn{1}{c|}{ILJ }  & \multicolumn{1}{c|}{ILJFH } & \multicolumn{1}{c|}{ILJ }  & \multicolumn{1}{c|}{ILJFH }             \\
\multicolumn{1}{|c|}{}        & \multicolumn{1}{c|}{}         & \multicolumn{1}{c|}{FIX}   & \multicolumn{1}{c|}{FIX}    & \multicolumn{1}{c|}{DEF}   & \multicolumn{1}{c|}{DEF}               \\ \hline
\multicolumn{1}{|c|}{250}     & \multicolumn{1}{c|}{106}         & \multicolumn{1}{c|}{146 (38\%)}   & \multicolumn{1}{r|}{86 (19\%)}     & \multicolumn{1}{c|}{559}   & \multicolumn{1}{c|}{382}     \\ \hline
\multicolumn{1}{|c|}{275}     & \multicolumn{1}{c|}{145}         & \multicolumn{1}{c|}{191 (32\%)}   & \multicolumn{1}{c|}{112 (23\%)}    & \multicolumn{1}{c|}{674}   & \multicolumn{1}{c|}{450}    \\ \hline
\multicolumn{1}{|c|}{300}     & \multicolumn{1}{c|}{189}         & \multicolumn{1}{c|}{224 (19\%)}   & \multicolumn{1}{c|}{151 (20\%)}    & \multicolumn{1}{c|}{728}   & \multicolumn{1}{c|}{563}     \\ \hline
\multicolumn{1}{|c|}{325}     & \multicolumn{1}{c|}{236}         & \multicolumn{1}{c|}{274 (16\%)}   & \multicolumn{1}{c|}{204 (14\%)}    & \multicolumn{1}{c|}{851}   & \multicolumn{1}{c|}{648}  \\ \hline
\multicolumn{1}{|c|}{350}     & \multicolumn{1}{c|}{285}         & \multicolumn{1}{c|}{360 (26\%)}   & \multicolumn{1}{c|}{245 (14\%)}   & \multicolumn{1}{c|}{915}   & \multicolumn{1}{c|}{736}   \\ \hline
\hline
\multicolumn{1}{|c|}{$E_0$}     & \multicolumn{1}{c|}{100 }         & \multicolumn{1}{c|}{96}   & \multicolumn{1}{c|}{109}   & \multicolumn{1}{c|}{62}   & \multicolumn{1}{c|}{76}   \\ \hline
\multicolumn{1}{|c|}{$\delta E_0$}     & \multicolumn{1}{c|}{$<$1}         & \multicolumn{1}{c|}{8}   & \multicolumn{1}{c|}{4}   & \multicolumn{1}{c|}{2}   & \multicolumn{1}{c|}{3}   \\ \hline
\end{tabular}
\caption[]{Permeances ($10^3$ GPU) for the transport of H$_2$ across GDY for various temperatures (K), as obtained from TDWP quantum calculations (ILJ force field and fixed membrane) and different MD simulations using either ILJ or ILJFH interactions and considering the membrane either fixed (FIX) or deformable (DEF). 
The relative errors of the ILJ and ILJFH MD permeances (fixed membrane) with respect to the quantum permeances is also indicated within parenthesis.
The last two rows of the Table present the activation energies and their errors ($E_0$ and $\delta E_0$, in meV) as obtained from the fits to Eq.~\ref{eq_arrh}, for the different simulations.
}
\label{Tab1}
\end{table}

From Fig.~\ref{Fig6} it can also be noticed that both permeances obtained from MD simulations grow with temperature at apparently the same rate than the quantum permeance.
To look at the temperature dependence in more detail, we assume an 
Arrhenius-type relationship for the flux ($J$) vs. 
temperature \cite{Cranford-Nanoscale-2012,Zhou-SciAdv-2020}:

\begin{equation}
    J = A\cdot exp(-\frac{E_0 N_A}{RT}) 
\label{eq_arrh}
\end{equation}

\noindent
where $N_A$ is the Avogadro number, $R$ is the molar gas constant, $A$ is a pre-exponential factor and $E_0$, an activation energy.
This is a reasonable assumption within the dilute gas limit, where each molecule experiences an independent energy barrier during the transport.
In the inset of Fig.~\ref{Fig6}, $1/RT$ is plotted against the logarithm of the flux $\log J$ for the three calculations with the fixed membrane, together with their fit to Eq.~\ref{eq_arrh} (see Fig.~S4 of the SI for an enlarged version of the inset).
There is a clearly linear dependence which validates our assumption and allows us
to estimate the activation energies of the different simulations considered. 
The lowest two rows of Table~\ref{Tab1} show these energies together with their errors ($\delta E_0$).
Notice that the activation energy for the quantum simulation ($E_0$=100 meV) coincides with the energy threshold of the quantum probabilities of Fig.~\ref{Fig-Probs}.
The classical ILJ and ILJFH activation energies (96 and 109 meV) have values that are fairly close to the quantum $E_0$, confirming our expectation that the dependence with temperature of the classical permeances is very similar to that of the quantum permeance.

\begin{figure}[h!]
\centering
\includegraphics[width=9.cm,angle=0.]{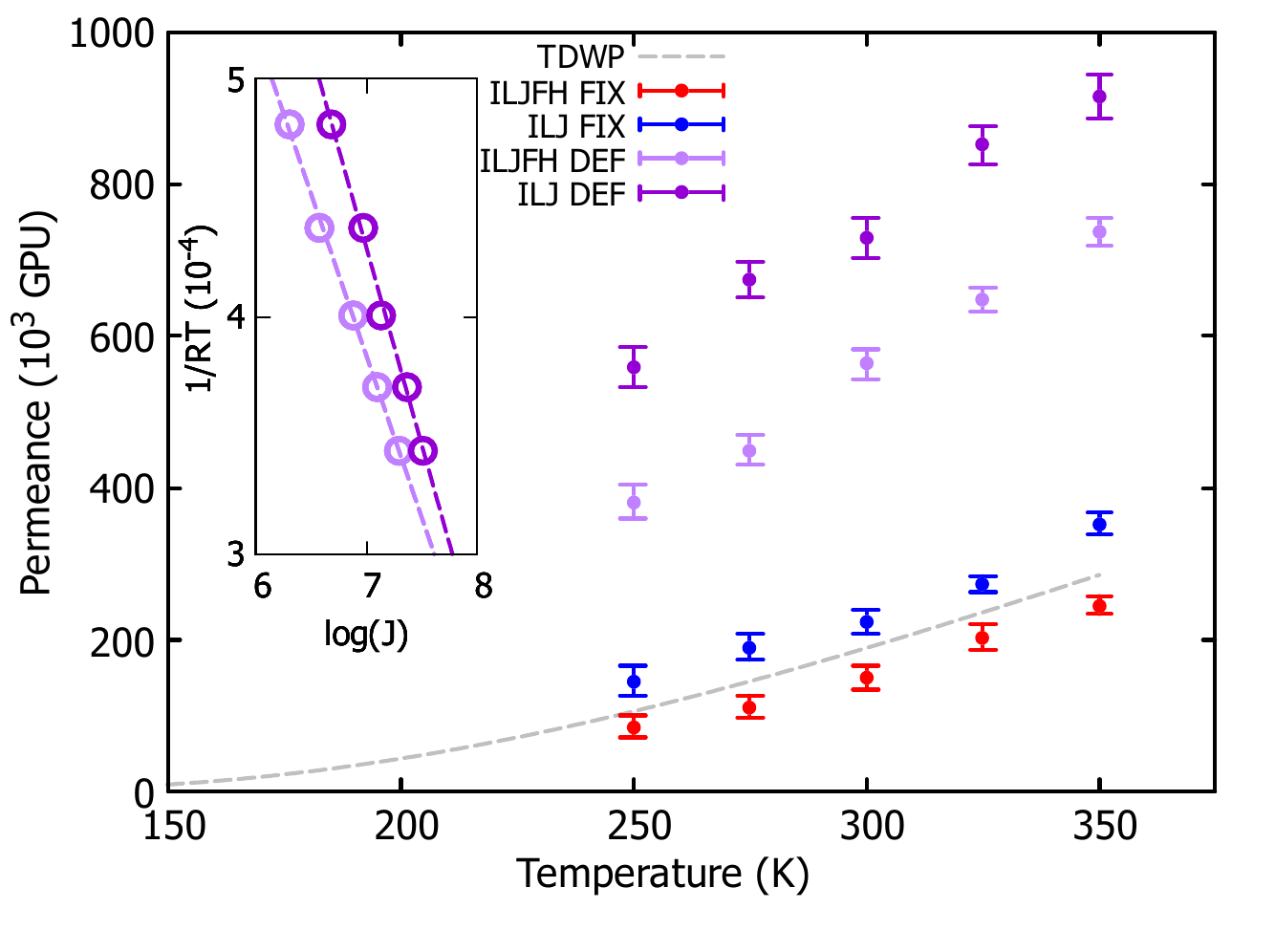}
\caption[] {Permeance (in $10^3$ GPU) of a deformable GDY membrane for the flow of H$_2$ molecules, as a function of temperature (K) and for ILJ+AIREBO (purple) and ILJFH+AIREBO (light purple) interactions. Classical (ILJ and ILJFH) and quantum permeances for the fixed membrane are also shown for reference. The inset shows $1/(RT)$ vs. the natural logarithm of the flux $J$ (in SI units) for the deformable membrane simulations (empty circles), while the dashed lines correspond to fits to Eq.~\ref{eq_arrh}.}
\label{Fig7}
\end{figure}

Within the MD framework, it is also possible to study the effect of the motion of the atoms of GDY (thermal and by collisions with the gas molecules).
In Fig.~\ref{Fig7} we report the classical permeances obtained using ILJ+AIREBO and ILJFH+AIREBO force fields, in the same temperature range studied in the case of the fixed membrane, while numerical values of these permeances are given in Table~\ref{Tab1}.
It can be seen that the permeances of deformable GDY are significantly larger than those of the fixed membrane, indicating that modeling the motion of this membrane is crucial in order to provide reliable predictions. 
For the ILJ (ILJFH) force field, they are larger than the permeance of the fixed GDY with a factor that ranges from 2.5 (3.0) for the highest temperature to 3.8 (4.4) for the lowest temperature.
Similar factors (ranging from 1.2 to 3.7) were found when comparing the fixed/deformable permeances in MD simulations of the transport of H$_2$, CO$_2$ and CH$_4$ mixtures across graphtriyine\cite{Azizi-SciRep-2021}.
We have also fitted the fluxes obtained for deformable GDY to Eq.~\ref{eq_arrh} in order to estimate activation energies (see inset of Fig.~\ref{Fig7} and Fig.~S4 of the SI) and the results are provided in the last two rows of Table~\ref{Tab1}.
The obtained activation energies (62 and 72 meV for ILJ and ILJFH) are considerably lower than those corresponding to the fixed membrane (about 100 meV).
This finding suggests that the energy barriers that the H$_2$ molecules experience when crossing the membrane are actually lower than in the case of the fixed membrane case, a feature that also explains the significant increase in the values of the permeances.

\begin{figure}[h!]
\centering
\includegraphics[width=9.2cm,angle=0.]{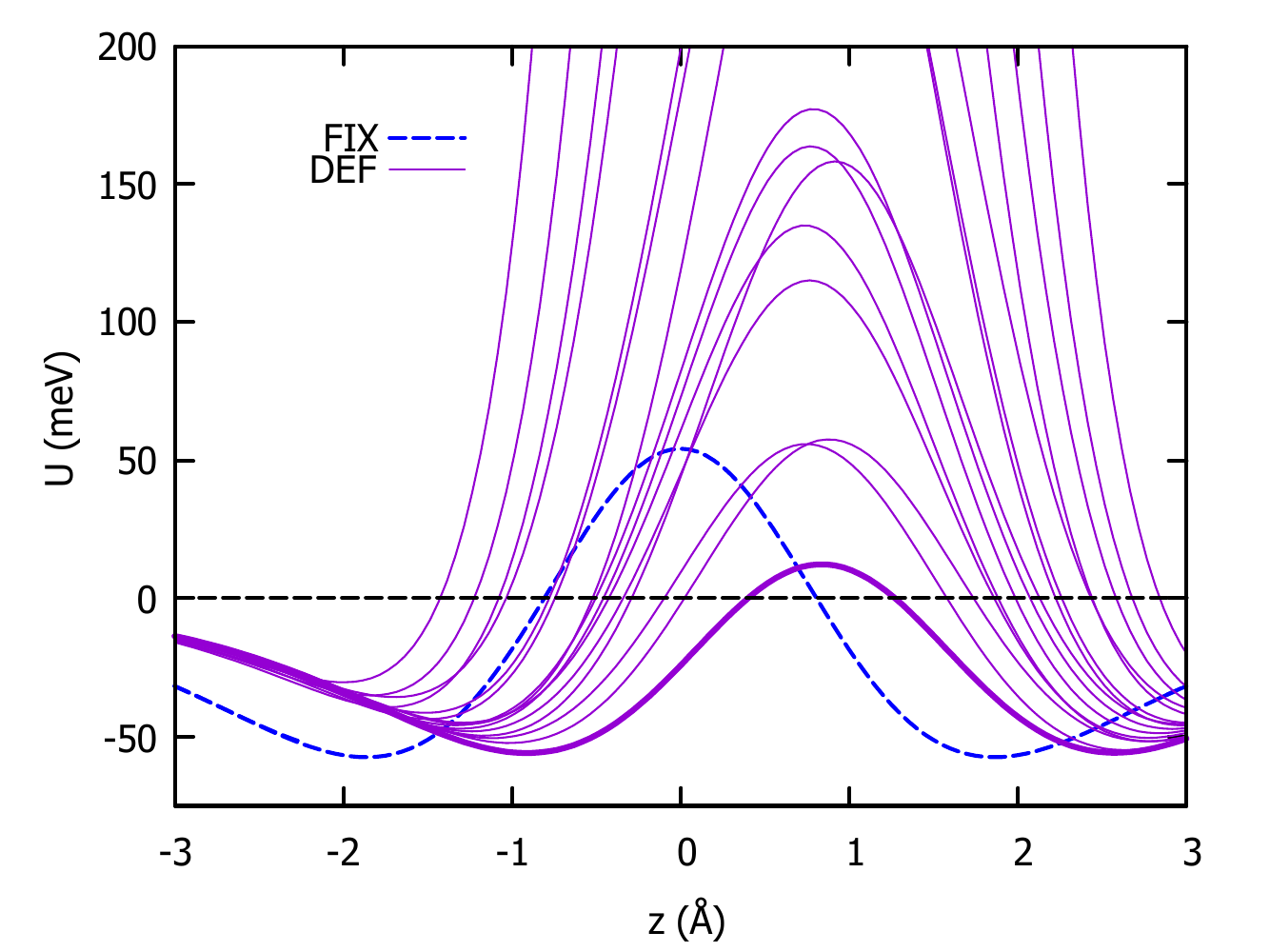}
\includegraphics[width=9.cm,angle=0.]{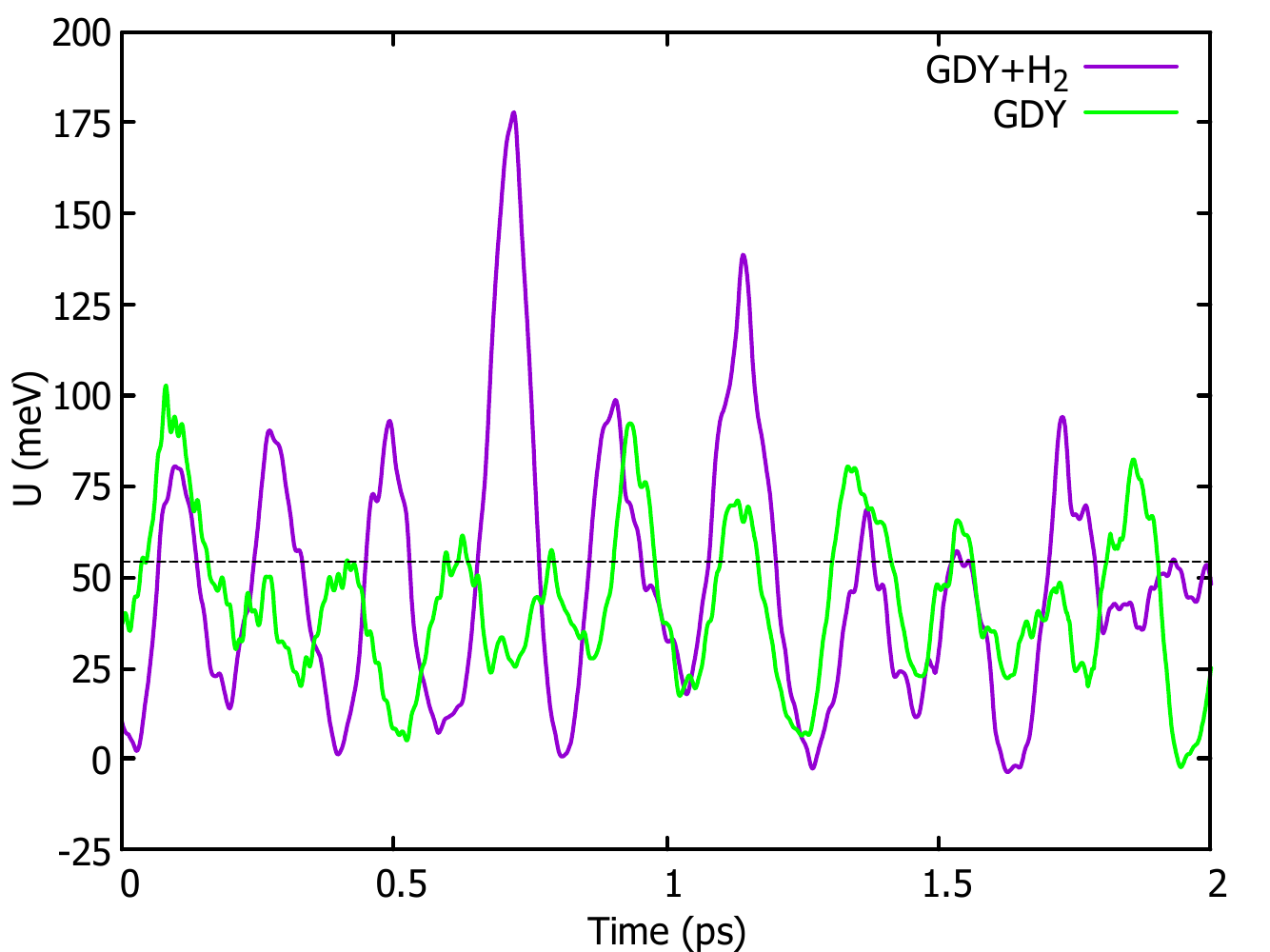}
\caption[] {Upper panel: Interaction energies between a GDY pore and a hydrogen molecule as a function of $z$ for a set of $(x_i,y_j)$ points around the center of the pore, at a given time  of a simulation with a deformable membrane (ILJ+AIREBO). The potential barrier for the crossing of H$_2$ through this pore is determined as the maximum value of the minimum energy path (highlighted line, leading to a barrier of 12 meV). The minimum energy path corresponding to a static pore (dashed blue line) is shown for comparison. Lower panel: Potential barrier obtained in this way as a function of time for the thermal motion (300 K) of the isolated membrane (green line) and of the membrane interacting with the hydrogen gas (violet). A previous equilibration period of 1 ns has been established.}
\label{Fig8}
\end{figure}

To understand in more detail the lower activation barriers found when the membrane motion is allowed, we have estimated the potential barriers of the deformable membrane as follows. 
At each timestep we have considered a region around the center of a pore $(x_0,y_0)$ and, for a set of $(x_i,y_j)$ values in this region, the interaction potential $U(z;x_i,y_j)$ between a H$_2$ and the pore is computed as a function of $z$.
From all the $(x_i,y_j)$ curves, the lowest energy path $U(z;x_{i_0},y_{j_0})$ is selected 
and the potential barrier is determined as the maximum value of this curve.
An example of the curves so obtained as well as the minimum energy path is depicted in the upper panel of Fig.~\ref{Fig8}.
In the example shown in Fig.~\ref{Fig8}, the potential barrier estimated in this way is 12 meV, considerably lower than the barrier of the static pore (48 meV).
This lower value is understood as a result of a widening of the pore as a consequence of the vibrations of the membrane.
Notice also the displacement of the pore's center of about 1 \AA  \, from its initial position ($z$=0), showing that this material is easily deformable (similar curves but shifted to about  -1 \AA  \, would have been obtained if the analysis involved a pore moving in the opposite direction, see Fig.~\ref{Fig_simbox}c).
The lower panel of Fig.~\ref{Fig8} presents the variation with time of the potential barrier  in two cases: the thermal motion of an isolated GDY membrane at 300 K and the motion of the membrane interacting with the hydrogen gas at the same temperature.
It can be seen that the potential barriers obtained in this way oscillate around the static potential barrier (horizontal black line) with large amplitudes and a periodicity of about 0.2 ps.
Interestingly, the amplitudes are larger in the case of the membrane interacting with the gas, the process becoming almost barrier-less at some points.
Thus, as the involved barriers drop below the static value about half the simulation time, 
much more molecules will be able to cross the membrane, those molecules with kinetic energies lower than the energies needed to cross the static barrier.

All the results in this work have been obtained with a GDY membrane containing 216 carbon atoms and size of about 3.2 $\times$ 2.8 nm$^2$, as indicated in the previous section.
In order  to validate the results of the deformable membrane simulations, we have performed further calculations with a larger membrane, specifically, of $8.2\times8.5$ nm$^2$ and containing 1620 carbon atoms ($5\times9$ unit cells).
The simulations were carried out with the same number of H$_2$ molecules (100), at 300 K and using the ILJ+AIREBO force field.
In the upper panel of Fig.~S6 we present a snapshot of an initial configuration of these simulations, whereas the lower panel shows the obtained average number of crossings as a function of time in comparison with the crossings corresponding to the small membrane area.
The resulting permeance of the larger membrane is $S_{1620} = (782 \pm 41)10^3$ GPU (for a total of 20 simulations), a value which compares quite well with the permeance from the small membrane simulations, $S_{216}= (728 \pm 26)10^3$ GPU (40 simulations).
We can conclude that the smaller GDY area allows us a reliable description of the effects of its motion on the transport of hydrogen molecules.

\section{Conclusion}

In this work we have studied the transport of hydrogen molecules through a graphdiyne membrane by means of both classical and quantum-mechanical methods. 
The quantum-mechanical treatment describes the approach of the molecule to a static membrane from any incidence direction in the three-dimensional space, therefore, this model allows us for the first time to compare with equivalent molecular dynamics simulations on almost equal terms.
In this way, previous quantum calculations of permeances have been extended to temperatures around 300 K, interestingly, finding that quantum effects (quantization of the transmission probabilities) are significant even at the relatively high kinetic energies involved.
As for the comparison with the molecular dynamics simulations, it is found that the classical permeances overestimate the quantum ones by about 16-38\% whereas introducing Feynman-Hibbs corrections to the employed force fields leads to a 14-23\% underestimation of the quantum permeances.
It is important to stress that  classical as well as quantum permeances exhibit a very similar dependence with temperature so that estimations of involved activation energies lead to very close values (between 96 to 109 meV).

Additionally, the role of the motion of the graphdiyne atoms in the transport dynamics has been examined within the classical approach, finding that the modeling of this motion is essential for a realistic study.
Indeed, by describing the carbon interactions with the AIREBO force field, we have obtained an enhancement of GDY permeances with respect to the static case (with enhancement factors of about 2.5-4.4), together with a notorious decrease of the activation energies (about 30\%).
To understand this result, we have estimated the potential barrier for the molecular crossing as the carbon atoms of a pore evolve with time due to thermal motion and collisions with the gas molecules.
It has been found that, due to the pore vibrations, the potential barrier significantly drops below the value of the static barrier for about half of the simulation time, so that a larger number of molecules approaching the pore will go through it, as the necessary kinetic energy is reduced. 
Finally, having seen that the two classical permeances obtained (with and without Feynmann-Hibbs corrections) "bracket" the quantum one for the fixed membrane case, the reported permeances within the deformable membrane model for these two kinds of classical simulations provide a hint for the values that might be expected for the quantum permeances of a mobile graphdiyne.

We expect that the approach followed in this study will be useful for related problems, such as isotopic separation or systems with multilayer membranes\cite{Campos-Martinez-ASS-2026}. A complementary and interesting approach would be the application of the so-called \textit{ab-initio} molecular dynamics simulations\cite{Mahnaee-S&I-2025,cahlik_acsnano_21,TRABADA2019260}.

\section*{Conflicts of interest}
There are no conflicts to declare.

\section*{Acknowledgments}
The work has been funded by Spanish Agencia Estatal de Investigaci{\'o}n  grant
PID2023-149406NB-I00 / AEI / 10.13039 / 501100011033.
M.R. thanks the Instituto de F{\'\i}sica Fundamental and Consejo Superior de Investigaciones Cient{\'\i}ficas for the award of a fellowship for initiation to research "JAE Intro ICU".
Allocation of computing time by CESGA (Spain) is also acknowledged.


\bibliography{Bib-Surf_Int}
\bibliographystyle{elsarticle-num-names.bst}

\end{document}